\documentclass[fleqn,usenatbib]{mnras}

\usepackage{newtxtext,newtxmath}

\usepackage[T1]{fontenc}
\usepackage{ae,aecompl}
\usepackage{datatool}
\usepackage{hyperref}
\usepackage{bm}

\usepackage{graphicx}	
\usepackage{amsmath}	
\newcommand\textlcsc[1]{\textsc{\MakeLowercase{#1}}}

\title[Dark matter wakes and phase spirals]{An ever-present {\it Gaia} snail shell triggered by a dark matter wake}

\author[R. J. J. Grand et al.]{\parbox[t]{\textwidth}{
Robert J. J. Grand$^{1,2,3}$\thanks{E-mail: r.j.grand@ljmu.ac.uk}, R{\"u}diger Pakmor$^4$, Francesca Fragkoudi$^5$, Facundo A. G{\'o}mez$^{6,7}$, Wilma Trick$^4$, Christine M. Simpson$^8$, Freeke van de Voort$^9$, Rebekka Bieri$^{4,10}$}
\vspace{10pt}
\\
$^1$Astrophysics Research Institute, Liverpool John Moores University, 146 Brownlow Hill, Liverpool, L3 5RF, UK\\
$^2$Instituto de Astrof\'isica de Canarias, Calle Vía L\'actea s/n, E-38205 La Laguna, Tenerife, Spain\\
$^3$Departamento de Astrof\'isica, Universidad de La Laguna, Av. del Astrof\'isico Francisco S\'anchez s/n, E-38206, La Laguna, Tenerife, Spain\\
$^4$Max-Planck-Institut f\"{u}r Astrophysik, Karl-Schwarzschild-Str. 1, 85748 Garching, Germany\\
$^5$Institute for Computational Cosmology, Department of Physics, Durham University, South Road, Durham, DH1 3LE, UK\\
$^6$Instituto de Investigaci\'on Multidisciplinar en Ciencia y Tecnolog\'ia, Universidad de La Serena, Ra\'ul Bitr\'an 1305, La Serena, Chile\\
$^7$Departamento de Astronom\'ia, Universidad de La Serena, Av. Juan Cisternas 1200 Norte, La Serena, Chile\\
$^8$Argonne Leadership Computing Facility, Argonne National Laboratory, Lemont, IL 60439, USA\\
$^9$Cardiff Hub for Astrophysics Research and Technology, School of Physics and Astronomy, Cardiff University, Queen’s Buildings, Cardiff CF24 3AA, UK\\
$^{10}$Center for Space and Habitability, University of Bern, Gesellschaftsstrasse 6 (G6) 3012 Bern, Switzerland
}

\date{Accepted XXX. Received YYY; in original form ZZZ}

\pubyear{2022}

\begin{document}
\label{firstpage}
\pagerange{\pageref{firstpage}--\pageref{lastpage}}
\maketitle

\begin{abstract}
We utilize a novel numerical technique to model star formation in cosmological simulations of galaxy formation - called \textlcsc{Superstars} - to simulate a Milky Way-like galaxy with $\gtrsim10^8$ star particles to study the formation and evolution of out-of-equilibrium stellar disc structures in a full cosmological setting. In the plane defined by the coordinate and velocity perpendicular to the mid-plane (vertical phase space, $\{Z,V_Z\}$), stars in Solar-like volumes at late times exhibit clear spirals qualitatively similar in shape and amplitude to the {\it Gaia} ``Snail shell'' phase spiral. We show that the phase spiral forms at a look back time of $\sim 6$ Gyr during the pericentric passage of a $\sim10^{10}$ $\rm M_{\odot}$ satellite on a polar orbit. This satellite stimulates the formation of a resonant wake in the dark matter halo while losing mass at a rate of $\sim0.5$-$1$ dex per orbit loop. The  peak magnitude of the wake-induced gravitational torque at the Solar radius is $\sim 8$ times that from the satellite, and triggers the formation of a disc warp that wraps up into a vertical phase spiral over time. As the wake decays, the phase spiral propagates several Gigayears to present-day and can be described as ``ever-present'' once stable disc evolution is established. These results suggest an alternative scenario to explain the {\it Gaia} phase spiral which does not rely on a perturbation from bar buckling or a recent direct hit from a satellite. 
\end{abstract}




\section{Introduction}

Recent large Galactic surveys such as {\it Gaia} \citep{Gaia+Brown+Vallenari+18} have now crystallized the idea that the Milky Way is in a state of dynamical disequilibrium. The Galactic disc(s) in particular harbours a great deal of structure indicative of this, including vertical asymmetries such as the warp and Monoceros Ring \citep[e.g.][]{2012MNRAS.423.3727G,Slater2014,Xu2015,Poggio2018,SchoenrichDehnen2018} and planar stellar moving groups \citep[see e.g.][]{Katz2018,Kawata2018,Antojaetal2018,Fragkoudi2019}. These features are thought to originate from rich dynamical phenomena ranging from external perturbations \citep[e.g.][]{2012ApJ...750L..41W,Xu2015,2017MNRAS.465.3446G,Antojaetal2018,Laporte2022} to internal processes from bars and spiral arms \citep[e.g.][]{2016MNRAS.461.3835M,Fragkoudi2019,Trick2019,Trick2021}. The detailed information now available from large Galactic surveys provide a unique opportunity to learn about these gravitational processes that have shaped the distribution of stars in our Galaxy. Given that gravity permeates the dark sector as well as the baryonic, it is also a window into the distribution of dark matter, and perhaps even its nature.

One of the most striking and recently discovered dynamical features discovered in the Galactic disc is the so-called {\it Gaia} ``phase spiral'' or ``Snail shell'' \citep{Antojaetal2018,Tianetal2018,LMJ19,Frankel2023}: a spiral pattern in the vertical phase plane ($Z, V_Z$) associated with oscillations perpendicular to the Galactic plane of nearby disc stars. This feature can be seen in both the density and planar velocity (either $V_{\phi}$ or $V_{R}$) of stars in this plane, and is indicative of the phase mixing of a group of stars initially clumped together in the phase plane. The spiral shape occurs because the vertical period of oscillation is an increasing function of amplitude (anharmonic motion), therefore stars with larger oscillations in the initial clump would take longer to traverse a phase space ellipse compared to those with smaller oscillations \citep[see ][for a thorough explanation]{Binney_Schoenrich2018}.

There has already been a surge in activity to try to understand the origin of the initial clump of stars. Several studies have linked the phase spiral to a perturbation from a dwarf satellite \citep[e.g.][]{Antojaetal2018,Binney_Schoenrich2018}. Such a perturbation can qualitatively reproduce the phase spiral as seen in $V_{\phi}$ or $V_{R}$ by generating correlated, coherent in-plane and vertical oscillations of stars entering the Solar neighbourhood from both the inner and outer disc; a clump of low-$V_{\phi}$, short vertical period inner disc stars shears into tighter spirals relative to a clump of high-$V_{\phi}$, long vertical period outer disc stars. An obvious candidate for the perturbing satellite is the Sagittarius dwarf galaxy (hereafter Sgr), which is thought to have undergone several close pericentric passages \citep{PB10,Ruiz-Laraetal2020} and induced gravitational perturbations on the disc \citep{2013MNRAS.429..159G}. Indeed, several idealised $N$-body simulations have shown that earlier passages of a more massive Sagittarius are able to qualitatively reproduce the phase spiral \citep[][]{LMJ19,BHTC2021,Huntetal2021}.

However, the explanation of the Sagittarius dwarf galaxy as the source of perturbation is not completely accepted. For example, \citet{Bennettetal2022} conclude that the inferred present day mass of the Sgr core remnant \citep[$\sim 3\times 10^8$ $\rm M_{\odot}$,][]{Vasliev_Belokurov2020} is too low to have produced the observed amplitude of the spiral on its own, although their experiments did not allow for significant mass loss over multiple passages of Sgr \citep[see][]{BHTC2021}. Nevertheless, alternative sources of perturbation such as spiral arms and/or a bar have gained traction in the literature: \citet{Khoperskovetal2019} advocates for a strong vertical perturbation caused by a buckling bar, \citet{Tremaine2023} discuss Gaussian noise from dark matter subhalos and Giant Molecular Clouds, whereas \citet{Darling_Widrow2019} discuss bending waves generated by an {\it ad hoc} vertical perturbation. It is worth noting that vertical perturbations developed by internal mechanisms typically show smaller amplitudes than those associated with external agents, such as a large satellite \citep[e.g.][]{2014MNRAS.440.2564F,2016MNRAS.461.3835M,2021ApJ...908...27G}. The recent work of \citet{Garcia-Condeetal2022} found from their cosmological simulation that several relatively light satellites appear to be connected with the genesis of a phase spiral-like features, which indicates that the situation is not fully understood and could be more complex than previously imagined.

Another class of perturbation arise from collective effects from resonant wakes generated in the dark matter halo by a passing satellite \citep[e.g.][]{1995ApJ...455L..31W, 1998MNRAS.299..499W, Vesperini-Weinberg2000}. Cosmological simulations have shown that these wakes can induce dynamical perturbations more than an order of magnitude larger than those from the satellite itself, and form galaxy-wide disc warps and corrugation patterns with features similar to that of the Milky Way's Monoceros Ring \citep{GWM15, 2021ApJ...908...27G}. Using tailored $N$-body simulations of the impact of a Sgr-like galaxy on an equilibrium stellar disc, \citet{Laporte2018,LMJ19} showed that the early pericentric passages of Sgr stimulate the growth of a dark matter wake and the appearance of phase spirals. However, at late times, the wake impact becomes negligible compared to the direct impact of Sgr. The latter ``resets'' the vertical disc structure into a phase spiral configuration likened to the Gaia Snail shell. Thus the nature of the perturbation from which the observed feature arises is attributed directly to Sgr itself.

Nearly all of the theoretical work discussed above that explicitly studies the phase spiral adopts either toy or idealised $N$-body models \citep[to the author's knowledge, the single exception is][]{Garcia-Condeetal2022}. These models do not include cosmologically-grown stellar discs (and therefore have no memory of stellar populations formed during past epochs) that could respond differently to perturbations relative to smooth equilibrium discs. Nor do they include the array of perturbations inherent to a cosmological setting, such as misaligned gas discs, the spectrum of subhaloes \& satellites expected for the $\Lambda$CDM paradigm, and a non-spherical dark matter halo. Cosmological zoom-in simulations model all of these processes, but their limited resolution typically precludes the study of delicate and detailed dynamical features like the phase spiral. For example, \citet{GWM15} globally characterized the vertical response of a galactic disc simulated on a fully cosmological context, but lacked the resolution to study its response in local Solar-like volumes. For cosmological simulations to match the detail provided by $\sim 10^8$ star particles now attained by idealised models \citep{BHTC2021,Huntetal2021}, tens of millions of cpu hours per simulation are required\footnote{Few cosmological simulations have attained such a high stellar particle resolution, namely the Justice League Mint Condition simulation \citep{Applebaum_etal2021} and a simulation from the Auriga project \citep{GMP21}.}. This substantial computational expense, which is mainly incurred by the hydrodynamic calculation involving large numbers of gas particles/cells, hinders the production of cosmological simulations capable of resolving detailed dynamical structures such as the snail shell.

In this paper, we employ a new technique for star formation in cosmological simulations - called \textlcsc{Superstars}, which significantly boosts the stellar resolution to $\gtrsim 10^8$ star particles without the need to increase the gas resolution. This approach yields significant advantages: it provides access to new dynamical scales for stars at a substantially reduced computational cost, and side-steps the most significant challenges to numerical convergence which are driven almost entirely by changes to gas resolution. We describe this technique in Section~\ref{sim}. In Section~\ref{results}, we study the nature and origin of dynamical features analogous to the {\it Gaia} ``Snail shell''. We show that the simulated disc develops a spiral structure in vertical phase space during the epoch of disc formation, and tie its origin to a dark matter halo wake. We show that this feature lasts until the present day, by which time it has decayed to an amplitude and shape quantitatively similar to the {\it Gaia} phase spiral. In Section~\ref{conclusions}, we summarise our conclusions and discuss our findings in the context of earlier work.

\section{Simulations}
\label{sim}
\subsection{The Auriga model}

The simulated galaxy presented in this paper is a re-simulation of one of the Milky Way-mass systems from the \textlcsc{Auriga} project \citep[][]{GGM17,GHF18}, specifically the halo presented in \citet{GMP21} (referred to as Au 6 in \textlcsc{Auriga} nomenclature). This halo has a mass of $M_{200} = 1.03 \times 10^{12} \rm M_{\odot}$ at redshift zero, where $M_{200}$ is defined as the mass contained inside the radius at which the mean enclosed mass density equals $200$ times the critical density of the universe. The parent dark matter only cosmological simulation has a comoving periodic box size 100 Mpc, and adopts the following parameters for the standard $\Lambda$CDM cosmology: $\Omega _m = 0.307$, $\Omega _b = 0.048$, $\Omega _{\Lambda} = 0.693$ and a Hubble constant of $H_0 = 100 h$ km s$^{-1}$ Mpc$^{-1}$, where $h = 0.6777$, taken from \citet{PC13}. At redshift 127 (the starting redshift), the resolution of the dark matter particles of the Lagrangian region from which this halo forms is increased and gas is added to create the initial conditions of the zoom simulation. At redshift zero, this high resolution region has a radius of the order $\sim 1$ Mpc. 

The simulation was performed with the magneto-hydrodynamic code \textlcsc{AREPO} \citep{Sp10,PSB15}, and the \textlcsc{AURIGA} galaxy formation model, which includes: primordial and metal line cooling; a uniform UV background that gradually increases to completion at $z=6$; a model for star formation that activates for gas densities larger than $0.1$ atoms $\rm cm^{-3}$ \citep{SH03}; magnetic fields \citep{PMS14,PGG17,PGP18}; gas accretion onto black holes and energetic feedback from AGN and supernovae type II  \citep[SNII, see][for more details]{VGS13,MPS14,GGM17}. Each star particle is treated as a single stellar population of given mass, age and metallicity. Stellar mass loss and metal enrichment from type Ia supernovae (SNIa) and Asymptotic Giant Branch (AGB) stars are modelled according to a delay time distribution, and metals from SNII are injected promptly. The \textlcsc{AURIGA} model has been shown to produce realistic spiral disc galaxies that are broadly consistent with a number of observations including star formation histories, stellar masses, sizes and rotation curves of Milky Way-mass galaxies \citep{GGM17}, the distribution of HI gas \citep{MGP16}, the stellar halo properties of local galaxies \citep{MGG19}, stellar disc warps \citep{GWG16}, the properties and abundance of galactic bars \citep{Fragkoudi+Grand+Pakmor+19,Fragkoudi2021} and bulges \citep{GMG19}, the properties of magnetic fields in nearby disc galaxies \citep{PGG17,PGP18}, and the luminosity function of satellite galaxies \citep{SGG17}. 

In the study of \citet{GMP21}, we presented the hitherto highest resolution cosmological hydrodynamic zoom simulation of a Milky Way-mass halo; the baryonic and dark matter mass resolution of the simulation is $\sim 800$ $\rm M_{\odot}$ and $ 6 \times 10^3$ $\rm M_{\odot}$, respectively. In the \textlcsc{AURIGA} nomenclature, this resolution is given the shorthand ``level 2''. Apart from the significant computational expense ($\sim15$ million CPU hours), this study highlighted two significant issues: i) a $\sim 30\%$ systematic increase in stellar mass of the main galaxy for each factor 8 increase in mass resolution \citep[see Table~2 of][]{GMP21}; ii) a break-down in the black hole centering algorithm at very high (level 2) gas resolution with negative consequences for disc formation. The former is a qualitatively generic problem for all hydrodynamic simulations, whereas the latter is a new obstacle. Both, however, are related to increases in gas resolution. This situation motivates a different approach to model stellar dynamics in cosmological simulations in which the mass resolution of collisionless components are enhanced relative to the gas.

\subsection{The Superstars method}

We adopt a newly developed method called \textlcsc{Superstars} which achieves both a very high stellar resolution and removes the issues described above, namely: i) the large computational cost; and ii) systematic changes with gas resolution. This method will be fully described in Pakmor et al. in prep, and the full suite of simulations will be presented in Fragkoudi et al. in prep. Here, we briefly summarise the essence of the method. Instead of forming a single star particle of a mass approximately equal to that of the gas cell from which it spawned, \textlcsc{Superstars} forms a group of lower-mass star particles instead. The birth positions are identical for each of the star particles in one group. Their velocities are set to the velocity of the parent gas cell plus a random isotropic component. The size of the isotropic component is drawn randomly from a Gaussian distribution with a width set by the minimum of the local sound speed and velocity dispersion of its neighbouring gas cells. We ensure that the total contributions of all random components of one group cancel to conserve total momentum in the simulation. The chemical evolution is handled in exactly the same manner as the original \textlcsc{AURIGA} simulations. The number of star particles formed per group is in principle arbitrary, but is naturally limited by the available computational resources. In the simulation discussed in this paper, we retain the level 4 gas resolution ($\sim 5 \times 10^4$ $\rm M_{\odot}$) and form 64 star particles per group and star-forming gas cell. This achieves the same stellar mass resolution ($\sim 800$ $\rm M_{\odot}$) as the simulation presented in \citet{GMP21} with the highly desirable benefits of much improved numerical convergence (including a well-behaved black hole centering algorithm) and a more than $10$ times reduction in the overall computational cost. As will be shown in Pakmor et al. in prep., the larger dark matter to stellar particle mass ratio that this technique entails does not enhance artificial scattering of particles found for lower resolution large cosmological volume simulations \citep{Ludlow2019,Ludlow2021}. 

In the context of the present study, \textlcsc{Superstars} resolves detailed Galactic structure of the kind recently observed by large surveys such as {\it Gaia} in the presence of an array of complex dynamical phenomena inherent to galaxy formation. This is complementary to studies based on toy models and idealised $N$-body simulations that make up the vast majority of the current literature on the subject.

\section{Results}
\label{results}

\subsection{Present-day phase spiral properties}
\label{results:1}

\begin{figure*}
\includegraphics[scale=0.68,trim={0 0 0 0}, clip]{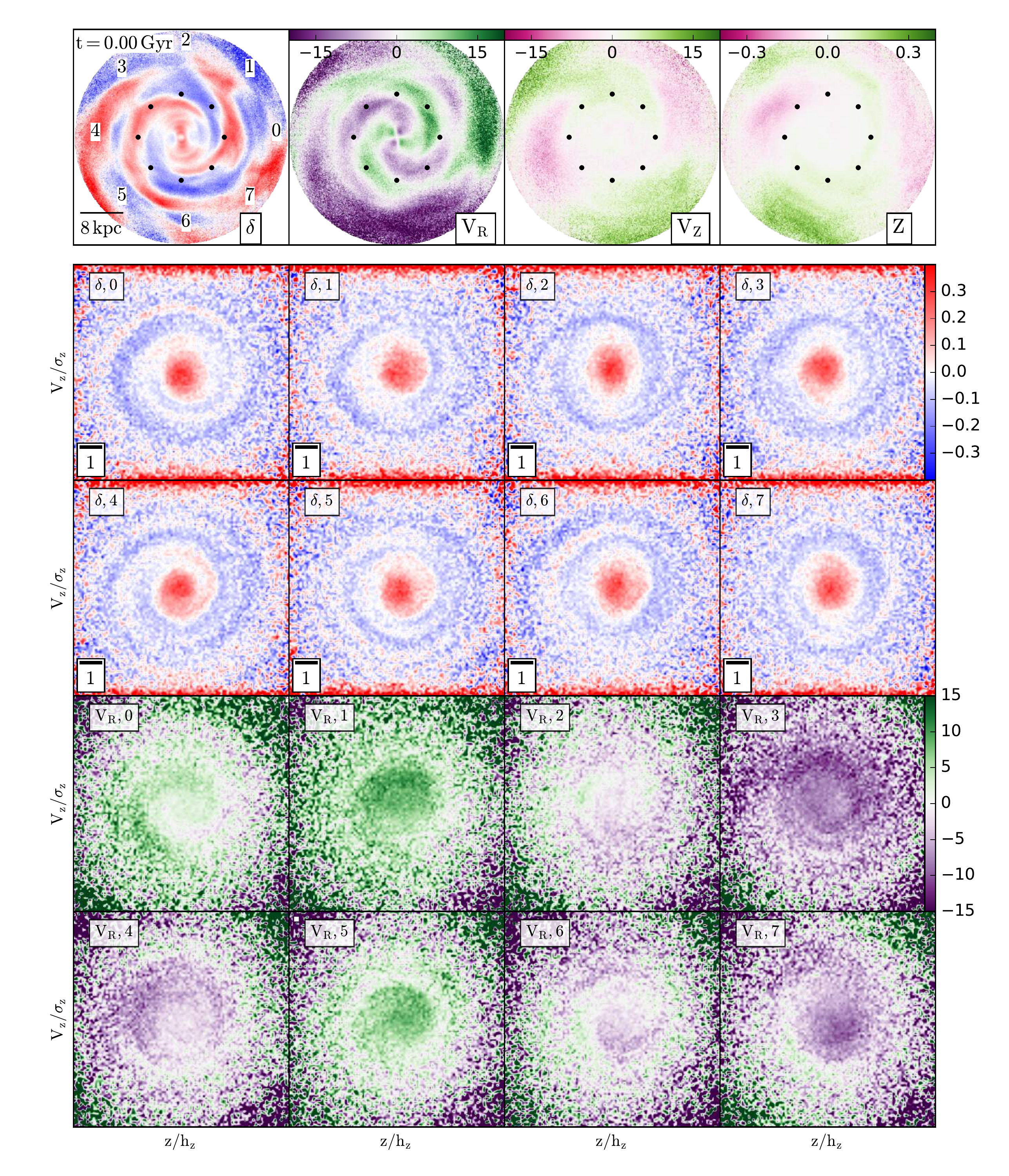}
\caption{{\it Top row:} 40 kpc $\times$ 40 kpc face-on projections of various quantities for star particles within 1 kpc height above/below the midplane at redshift zero: the azimuthal over-density (leftmost panel); the mean radial velocity ($\rm km\, s^{-1}$, second panel); the mean vertical velocity ($\rm km\, s^{-1}$, third panel); and the mean vertical height (kpc, fourth panel). Black symbols mark the positions of 8 Solar-like positions placed equidistant along a ring of 8 kpc radius. {\it Second \& third rows:} the over-density of star particles located within 3 kpc of each Solar-like position in the dimensionless vertical phase plane (see text for details). {\it Fourth \& fifth rows:} as above, but coloured according to radial velocity ($\rm km\, s^{-1}$). The phase-space spiral is visible at all eight Solar-like locations. A high-cadence (5 Myr time resolution) animation of this figure in the co-rotating frame (at $R=8$ kpc) can be viewed at \url{https://wwwmpa.mpa-garching.mpg.de/auriga/movies/multi_halo_6_sf64.mp4}. Fig.~\ref{multi:lowres} in Appendix~\ref{appa2} is the equivalent figure for a ``level 4'' simulation, which demonstrates that phase spirals are not resolved at that resolution.}
\label{multi:now}
\end{figure*}

\begin{figure}
\includegraphics[scale=0.6,trim={0 5cm 0 0}, clip]{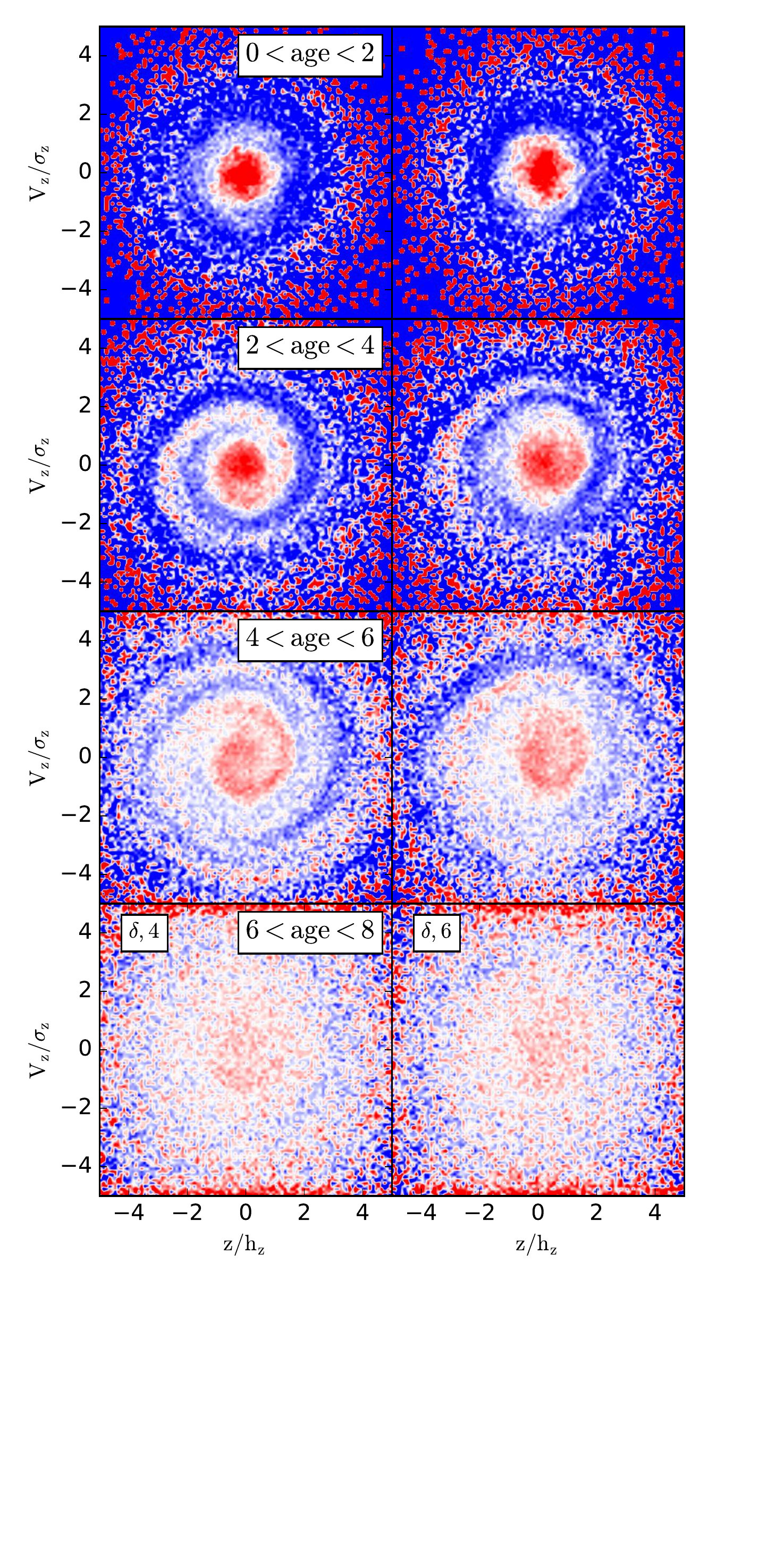}
\caption{The vertical phase plane for of star particles of different ages within 3 kpc spheres at two different Solar-like positions: volumes 4 (left column) and 6 (right column) as indicated in Fig.~\ref{multi:now}. From top to bottom, the age groups are: younger than 2 Gyr; between 2 and 4 Gyr old; between 4 and 6 Gyr old; and between 6 and 8 Gyr old. The phase spiral is clearest in the two intermediate age groups (second and third rows), however very faint signs are visible in the oldest age group.}
\label{spiral:age}
\end{figure}

\begin{figure*}
\includegraphics[scale=0.7,trim={0 0 0 0}, clip]{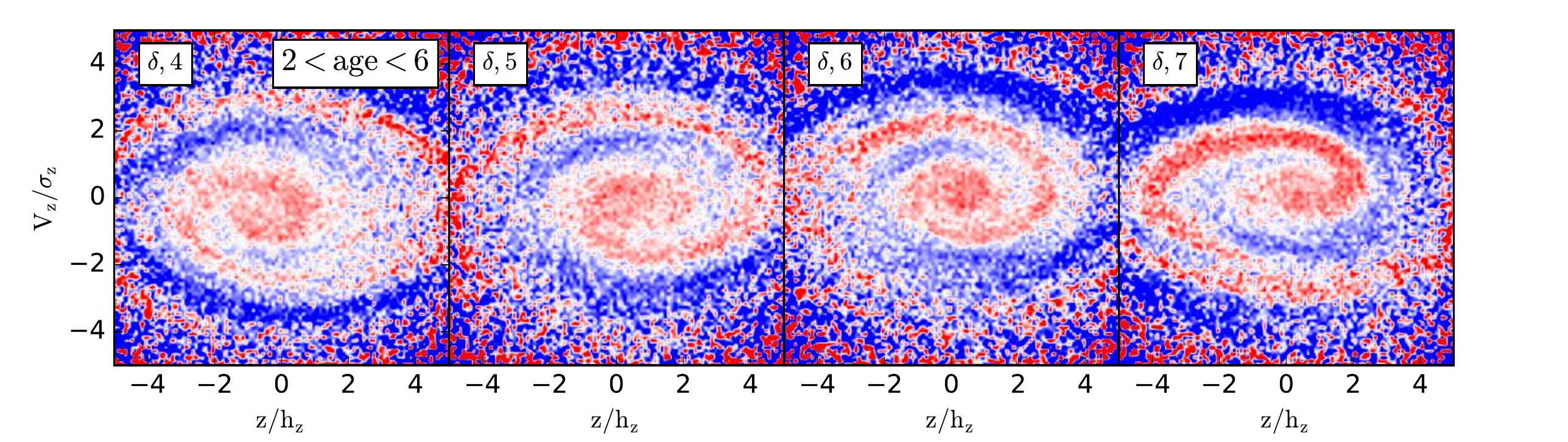}
\caption{Similar to Fig.~\ref{spiral:age}, but for star particles between 2 and 6 Gyr old within 3 kpc spheres centred on positions along a ring of radius 14 kpc.}
\label{spiral:age2}
\end{figure*}

In this section, we present the properties of the disc and vertical phase spiral at redshift zero, with reference to the {\it Gaia} snail shell where appropriate\footnote{We note that the purpose of this section is not to make detailed quantitative comparisons with observations, but rather place our results in the context of the literature.}. The top row of Figure~\ref{multi:now} shows a face-on view of the present-day stellar disc colour-coded according to various properties: the azimuthal stellar over-density in the disc plane (left panel); the mean radial velocity (second panel); the mean vertical velocity (third panel); and the mean vertical height (fourth panel). Here, we note the presence of a weak bar and clear spiral structure stretching from the ends of the bar into the outer disc. In a similar morphological pattern, the mean radial velocity betrays streaming motions of magnitude $\sim 10$ $\rm km\, s^{-1}$ correlated with the bar/spiral over-densities. A mild corrugation pattern is evident in the third and fourth panels of the top row of Fig.~\ref{multi:now} as oscillations in $Z$ and $V_Z$ along radial ``spokes'' of constant azimuth; the pattern appears to span from the Solar circle to the edge of the galaxy. 

The second and third rows of Fig.~\ref{multi:now} show the over-density of the $V_Z$-$Z$ distribution of star particles within 3 kpc of 8 Solar-like positions (spread equidistant in azimuth along a cylindrical radius of 8 kpc in the disc mid-plane), where each coordinate is normalised such that they are dimensionless \citep[as done in][for example]{Huntetal2021}. For the normalisation factors, we calculate, for each snapshot, the standard deviation of the vertical position\footnote{Note that this is not the same as the scale height, $Z_0$, of a fitted density profile typically used to measure the thickness of discs, such as the $\rm{sech}^2(Z/Z_0)$ profile. For this simulated galaxy, the thin and thick disc scale heights are: $Z_{\rm 0, thin}= 363$ pc and $Z_{\rm 0, thick}= 1107$ pc, respectively. These values are consistent with current estimates for the Milky Way's thin and thick discs \citep[see][and references therein]{BHG16}.} and velocity ($h_Z$ and $\sigma_Z$, respectively) of star particles younger than 3 Gyr in a Solar annulus of width and height equal to $2$~kpc. This selection ensures we calculate a reasonable normalisation for disc stars at Solar-like positions. At the present day, the values are: $h_Z=530$ pc and $\sigma _Z=23$ $\rm km\,s^{-1}$, respectively. To calculate the over-density value of each pixel in this surface of section, we first smooth the raw distribution with a 2D Gaussian kernel of width equal to 0.5 for both coordinates to yield a mean density map. We then divide the original unsmoothed map by the mean density map and subtract 1 from each pixel, such that pixels with a positive (negative) value are over-(under-)densities. Phase spiral structures are present at each Solar-like location and exhibit a variation in the detailed morphological appearance and amplitude: at some Solar-like positions (e.g, position 0), we see clear phase spirals\footnote{Note the Archimedean shape of these spirals ($r=a\theta$), as opposed to logarithmic spirals ($r=a e^{\theta \cot b}$) typically discussed in the context of galactic spiral arms \citep{BT08}.} with amplitudes of roughly $0.1$ in the region of phase space where they are clearly visible (typically at dimensionless radii of 0.5-2.5), which is also true without normalising $Z$ and $V_Z$ by $h_Z$ and $\sigma _Z$, respectively (see Fig.~\ref{multi:realdims}). These amplitudes are within $\sim 0.05$ of what is reported for the {\it Gaia} snail shell by \citet{LMJ19} from {\it Gaia} DR2 and by \citet{Hunt2022,Antoja2023} for {\it Gaia} DR3. However, just as in idealised models \citep[e.g.][]{LMJ19,Hunt2022,Bennettetal2022}, the dimensional phase spiral in this cosmological simulation extends farther than the {\it Gaia} snail shell (see Appendix~\ref{appa2} and Fig.~\ref{multi:realdims}).

The fourth and fifth rows of Fig.~\ref{multi:now} are similar to the second and third rows but shows the mean galactocentric radial velocity of star particles in each pixel. First, we note that the average radial velocity in each vertical phase space varies across the disc; the baseline of the radial velocity fluctuations in regions in which the mean radial streaming velocity is outward (particularly positions 1 and 5; see second panel of the top row) is shifted such that the minimum velocity is only just below 0, and {\it vice versa} for positions 3 and 4. Note that the mean radial velocity in each Solar-like position oscillates between inward and outward as the azimuth changes, which appears correlated to spiral arm over-densities. Taking into account this variation among different Solar-like positions, we deduce that the radial velocity amplitude of the phase spiral to be approximately 10 $\rm km \, s^{-1}$, which is comparable to that of the observed phase spiral \citep[e.g.][]{LMJ19}.

\begin{figure*}
\includegraphics[scale=1.,trim={2cm 15cm 2cm 2cm}, clip]{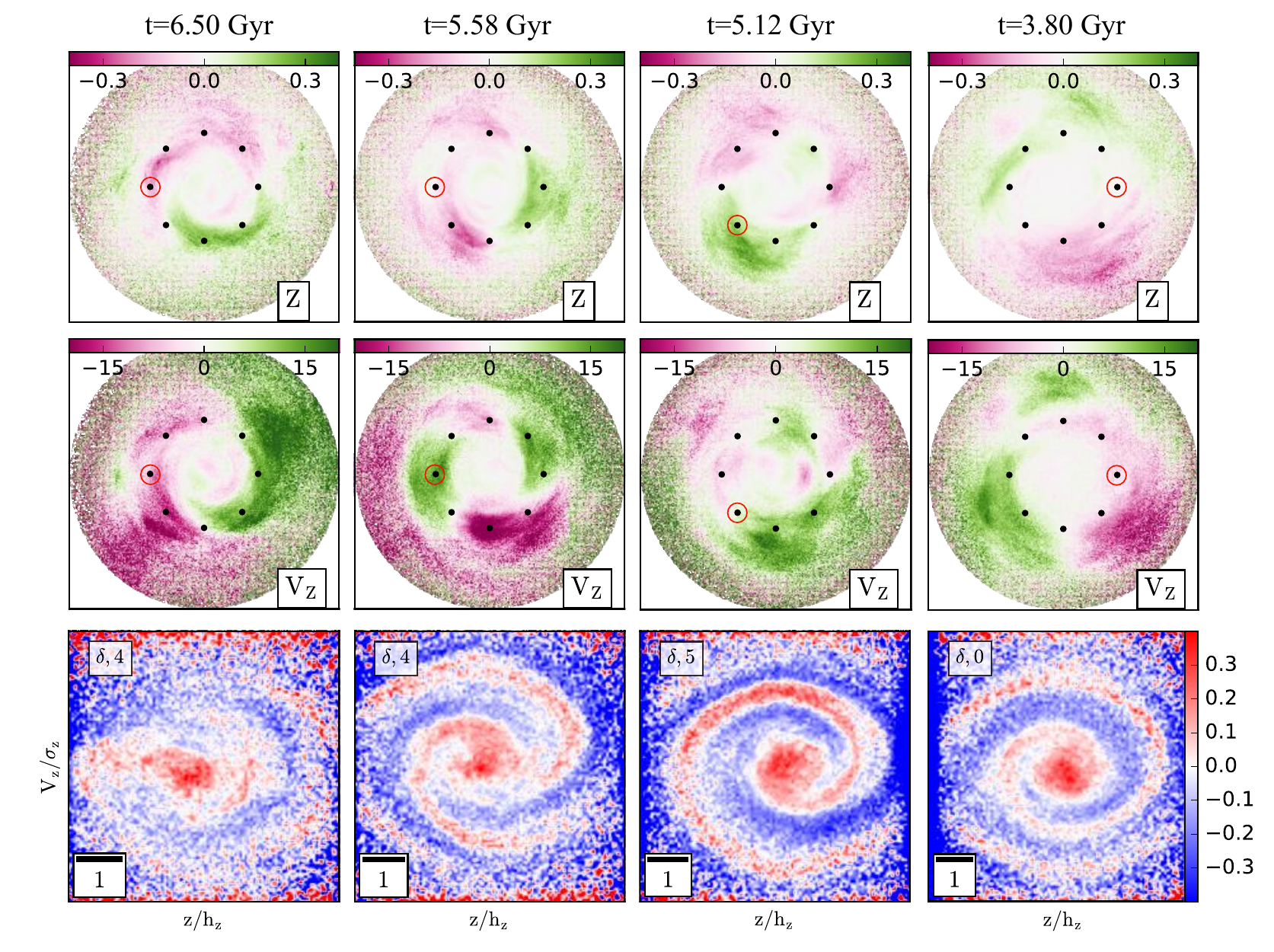}
\caption{Time evolution of the warp/corrugation and snail shell. {\it First and second rows:} the mean vertical height and vertical velocity of stars for face-on apertures of 40 kpc $\times$ 40 kpc at a series of times. Colour bars indicate the colour scale in units of kpc and $\rm km\,s^{-1}$, respectively. Black symbols mark the locations of 8 Solar-like positions. {\it Third row:} the phase spiral over-density at a Solar-like position (indicated by the label in the top-left of each panel, and by the red circle in the first and second rows), selected to clearly show the structure at each time. The size of the phase plane is indicated by the scale bar in the lower-left corner of each panel; note that the aperture becomes larger from left to right as the normalisation factors ($h_Z$, $\sigma _Z$) decrease with time owing to upside-down formation \citep[e.g.][]{GSG16}. These snapshots illustrate: the formation of a global warp pattern (first column) that quickly starts to wind up (first two rows). As shown in the last row, a short-lived ($\sim 100$ Myr) two-armed phase spiral emerges shortly after the onset of the warp (second column). Subsequently, a clear one-armed phase spiral develops (third column) and proceeds to wind-up and decrease in strength over time (fourth column). We remind the interested reader of the animation \url{https://wwwmpa.mpa-garching.mpg.de/auriga/movies/multi_halo_6_sf64.mp4} which shows the evolution of the phase spiral at each Solar-like position.}
\label{collage}
\end{figure*}

Several observational studies have dissected the vertical phase spiral as a function of age. One of the first such studies is \citet{Tianetal2018}, who analysed LAMOST and {\it Gaia} data and showed that the phase spiral is present among groups of coeval stellar populations younger than 6 Gyr, except perhaps for the very youngest stars (that formed less than 500 Myr ago) which may not exhibit a clear spiral. However, the subsequent studies of \citet{LMJ19} and \citet{Bland-Hawthorn2019} used isochrone ages derived by \citet{Sanders+Das18} and GALAH age estimates, respectively, to show the spiral to be present for all ages. As these studies discuss, this sort of dissection may help date the putative perturbation. To explore this idea in our simulation, we show in Fig.~\ref{spiral:age} the vertical phase space over-density for separate coeval stellar populations at two of the Solar-like positions shown in Fig.~\ref{multi:now} (positions 4 and 6; separated in azimuth by 90 degrees). Star particles that formed between 2 and 6 Gyr ago (shown in the second and third rows of Fig.~\ref{spiral:age}) show the strongest phase spirals. The oldest age group shows (at most) very faint signs of a phase spiral, likely because this population is kinematically hotter than the younger populations and therefore does not respond as coherently to dynamical perturbations as the latter. This is consistent with the work of \citet{GWM15}, who showed that satellite perturbations excited global warp and corrugation structures which were strongest in the youngest stellar populations. However, a lack of signal in stars older than 6 Gyr and younger than 2 Gyr is in contrast with the observational data presented by \citet{LMJ19} and \citet{Bland-Hawthorn2019}. We will discuss these results further in Section~\ref{conclusions}.

In Fig.~\ref{spiral:age2}, we show the vertical phase plane for star particles within 3 kpc spheres centred on positions along a ring of radius 14 kpc (but use the same normalisation factors as for the Solar-like positions in order to compare their relative shapes). The same age trends of the phase spirals described above for Solar-like positions hold also for the outer disc, therefore we show only star particles aged between 2 and 6 Gyr old in order to highlight the outer disc phase spiral properties clearly. We note two key differences compared to the spirals at Solar-like positions: i) the spirals are compressed along the vertical velocity axis relative to the vertical height axis, reflecting the lower vertical restoring force of the lower surface density outer disc; ii) the spirals are more loosely wound (fewer wraps) owing to the lower vertical frequencies and hence longer dynamical timescales of stars in the outer disc. These trends are consistent with those found in {\it Gaia} DR2 data and other simulations \citep[see e.g.,][]{LMJ19,Garcia-Condeetal2022}. Note also that these positions (4 - 7) span 135 degrees in azimuth (see the top-left panel of Fig.~\ref{multi:now}) and cover a downward and upward moving section of the corrugation pattern (see the third and fourth panels of the top row of Fig.~\ref{multi:now}). This translates to spirals in the $Z$-$V_Z$ plane that move through approximately 180 degrees from position 4 to position 7, and provides a flavour of the kind of variation that could be present in future observations covering broader swathes of the disc.

\subsection{The evolution of the phase spiral}
\label{results:evo}

Having shown that the morphology and strength of our simulated present-day phase spiral is similar to the Milky Way's snail shell, we now focus on its formation and evolution. The top two rows of Fig.~\ref{collage} shows a $\sim 3$ Gyr time sequence of the vertical height and velocity maps of the disc viewed face-on. The bottom row shows, for each snapshot, the vertical phase space over-density of star particles in a 3 kpc volume centred at a Solar-like location (indicated in each panel and selected to show the phase spiral particularly clearly). The first column shows this information for the snapshot $t_{\rm lookback}=6.5$ Gyr: the vertical height and velocity maps show a clear warp which stretches from the central disc to radii beyond the Solar-like positions (again marked by the black symbols). This warp manifests as an off-centre over-density in $Z$-$V_Z$ space (lower-left panel). We will show in Section~\ref{sec3:nature} that this occurs immediately after the close pericentric passage of a satellite galaxy of infall mass of $\sim 10^{10}$ $\rm M_{\odot}$. Over time, the disc warp winds into a corrugation pattern, the precise morphology of which evolves according to the radial dependence of vertical frequencies of stellar orbits and the relative strength of torques from the inner disc and outer halo \citep[see][for detailed explanations]{Briggs1990,ShenSellwood2006,GWG16}. In the vertical phase plane at $t=5.58$ Gyr, the distribution becomes more wrapped and a two-armed phase spiral is seen in $Z$-$V_Z$ space at the Solar location 4, which is in closest proximity to the disc crossing point of a satellite (we return to this in Section~\ref{conclusions}). However, this two-armed phase spiral is a local feature and lasts only approximately $100$Myr. The lower-third panel ($\sim50$~Myr later) and lower-fourth panel (a further $\sim1.3$~Gyr later) show clear one-armed vertical phase spirals: as phase mixing proceeds, the spiral becomes more tightly wound (characterised by more wraps) owing to the anharmonic motion discussed earlier. This time sequence clearly connects the phase spiral to an initially large-scale bi-symmetric warp, and thus supports the global nature of its origin \citep[see][for an idealised case]{LMJ19}.

\begin{figure}
\includegraphics[width=\columnwidth,trim={0 0 0.4cm 0.5cm}, clip]{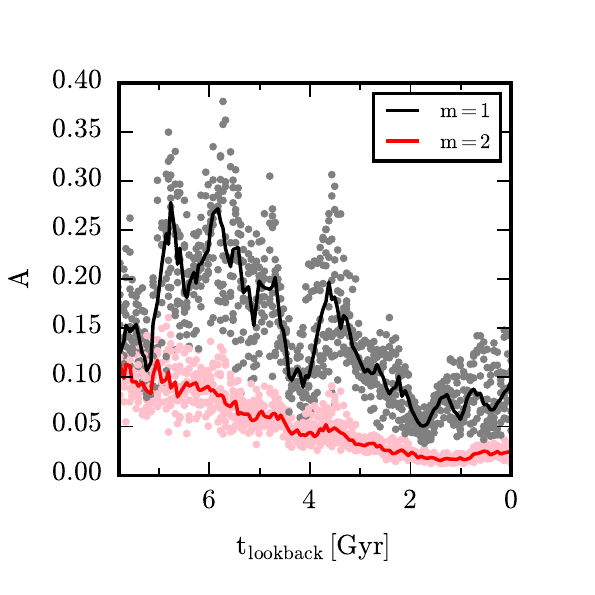}\\
\caption{The amplitude of the $m=1$ mode (black/grey) and $m=2$ mode (red/pink) phase spiral over-density (see Section~\ref{results:evo} for details) at each of the 8 Solar-like positions (dots) as a function of look back time. The medians of each mode amplitude across all positions are shown by the curves. The $m=1$ mode peaks in amplitude at $t_{\rm lookback}\sim 6$-$7$ Gyr, then proceeds to gradually decay to its present day shape and strength depicted in Fig.~\ref{multi:now}. The $m=2$ mode is weaker than that of the $m=1$ mode at all times.}
\label{psamp}
\end{figure}

To quantify the evolution of the phase spiral strength, we calculate the amplitude of the $m=1$ and $m=2$ terms of a discrete Fourier transform of the particles in the vertical phase plane:

\begin{equation}
    A= \sqrt{W_c^2 + W_s^2},
\end{equation}
where 

\begin{equation}
   W_c = \sum _i^N \cos{m\theta _i} / N; \\
   W_s = \sum _i^N \sin{m\theta _i} / N, 
\end{equation}
and $N$ is the number of particles, and $\theta _i$ is the angular coordinate of the $i$-th particle in this plane. The phase spiral appears well-resolved in the dimensionless distance range of 0.5 and 2.5, therefore we consider particles within this region only in order to avoid spurious measurement effects. We bin star particles into 10 equally spaced bins in dimensionless distance, calculate the amplitude in each bin, and take the median amplitude across all bins for each Solar position\footnote{The values obtained for the median amplitude at a given time and Solar-like position do not depend heavily on bin size.}. The evolution of the amplitudes of the $m=1$ mode and the $m=2$ mode are shown in Fig.~\ref{psamp}: individual Solar-like positions are represented by circles and the medians of these values for each time are shown by solid curves. The $m=1$ mode emerges at early times as the disc grows in earnest at $t_{\rm lookback}\sim 6$-$7$~Gyr, followed by a gradual decay lasting several billion years before reaching a present day amplitude of $\sim 0.1 \pm 0.05$. The $m=2$ mode amplitude is much weaker at all times, and only attains $A \gtrsim 0.1$ at early times for brief periods at a subset of locations.

\begin{figure}
\includegraphics[width=\columnwidth,trim={0 0.5cm 0.4cm 1cm}, clip]{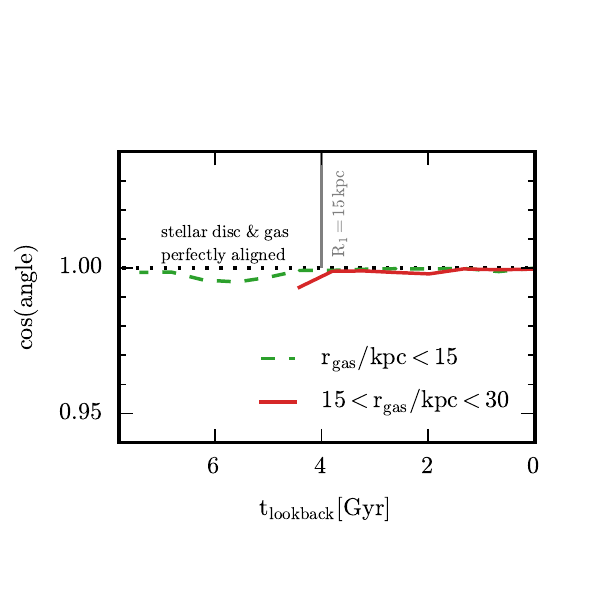}
\caption{Evolution of the cosine of the angle between the minor axis of the stellar disc and of the cold gas in different volumes indicated in the legend. The dotted black line marks the value corresponding to perfect alignment. The time at which the disc attains a size of 15 kpc is marked by a vertical grey line; the stellar disc is smaller than 15 kpc for times earlier than 4 Gyr look back time. Therefore, the gas within and around the disc is never more than 5 degrees out of alignment with the stellar disc: cold gas does not impart a significant torque on the disc.}
\label{gasangle}
\end{figure}

\subsection{The nature of the perturbation}
\label{sec3:nature}

\begin{figure*}
\includegraphics[scale=1.72,trim={0 0 0.7cm 0.5cm}, clip]{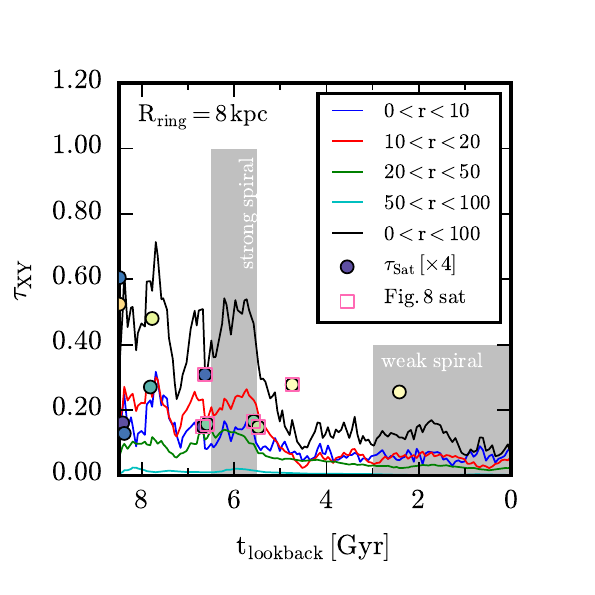}
\includegraphics[scale=1.72,trim={1.1cm 0 0 0.5cm}, clip]{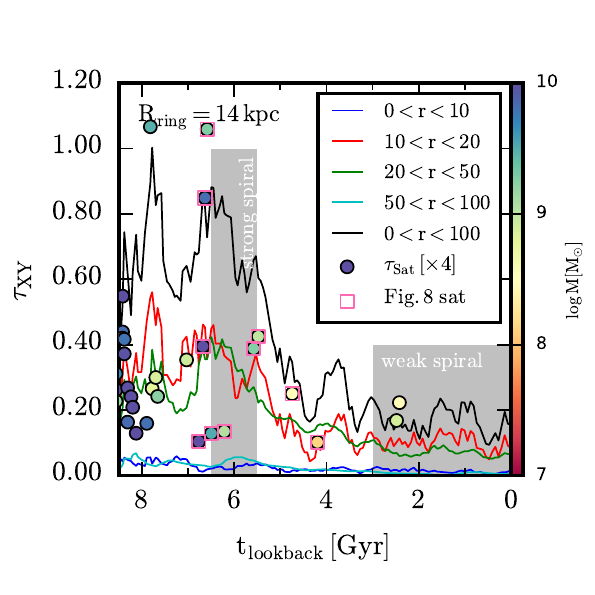}
\caption{Evolution of the magnitude of the plane-parallel components of torque, $\tau _{\rm XY}$, exerted by dark matter particles on an 8 kpc (left) and 14 kpc (right) ring of evenly distributed test particles oriented along the disc plane (see text for details). The different coloured curves indicate the torque associated with spherical shells of dark matter particles, as indicated in the legend. The circles show the torque exerted by satellite galaxies (multiplied by a factor of 4 to aid visual comparison) and are coloured according to their total mass. For clarity, only points with $\tau _{\rm XY} > 0.1$ are shown. The pink squares highlight the torque from a particular satellite (see Fig.~\ref{sat} and text for details). Note the green point at $t_{\rm lookback}\sim 6.5$ with $\tau _{\rm XY} \sim 1$ (right panel) and $\tau _{\rm XY} \sim 0.18$ (left panel) when the satellite was $\sim 5\times 10^9$ $\rm M_{\odot}$, which occurs of order 100 Myr prior to the peak torque from the dark matter halo inside 50 kpc: the latter torque is about 8 times larger than that of the former for $\rm R_{ring}=8$ kpc; and a factor of about 4 larger for $\rm R_{ring}=14$ kpc. The torque on each ring is dominated by the dark matter halo inside 50 kpc at all times. To guide the eye, grey shaded regions mark approximately the time periods when the phase spiral is particularly strong and weak (see Fig.~\ref{psamp}), which correlate with the amplitude of the dark matter torque.}
\label{dm:torq}
\end{figure*}

To understand the nature of the perturbation causing the phase spirals, we analyse the effects of possible perturbing sources, such as misaligned cold gas discs, dark matter, and satellites/subhaloes. As discussed earlier, the last of these has been extensively studied in the context of idealised simulations of the Sgr dwarf impact on a stellar disc. Dark matter halo torques (through, for example, dark matter wakes) has been found to have a significant impact on the stellar disc; it has been shown to be a key mechanism in the formation of galactic warps and corrugation patterns \citep[e.g.][]{GWM15,2021ApJ...908...27G,Laporte2018}. In addition, prior cosmological simulations \citep[e.g.][]{SWS09} have shown that a misalignment between cold gas and the stellar disc can have a significant dynamical (even destructive) impact on the stellar disc. However, with the exception of the recent study of \citet{Garcia-Condeetal2022}, cosmological simulations have not explicitly resolved/studied ``Snail shell''-like features, and therefore their connection to the aforementioned phenomena is unclear. In this section, we study the impact of each of these perturbing sources and isolate the main driver behind the phase spiral in our simulation.

First, we focus on the orientation of cold gas with respect to the disc as it grows over time. Fig.~\ref{gasangle} shows the evolution of the cosine of the angle between the minor axis of the stellar disc and that of two different volumes of cold star-forming gas: that contained within 15 kpc, and that found at radii between 15 and 30 kpc. At times earlier than $4$ Gyr, the edge of the stellar disc\footnote{Here we define the edge of the disc, $\rm R_1$, as the radius at which the surface mass density falls to $1 \rm M_{\odot} \, pc^{-2}$; the surface density drops off rapidly for larger radii.} is smaller than 15 kpc. Up to this time, the cold gas inside 15 kpc is never more than 5 degrees out of alignment with the stellar disc. At $\rm t_{lookback}=4$ Gyr, the stellar disc reaches a size of 15 kpc and continues to grow as it sustains near complete alignment with the cold gas within 30 kpc for the remainder of the evolution. The near-perfect alignment between the cold gas disc and the stellar disc does not produce a significant large-scale torque, therefore we conclude that misaligned gas is not a driver of the phase spiral in this simulation.

\begin{figure}
\includegraphics[width=\columnwidth,trim={0 1.7cm 0 1.5cm}, clip]{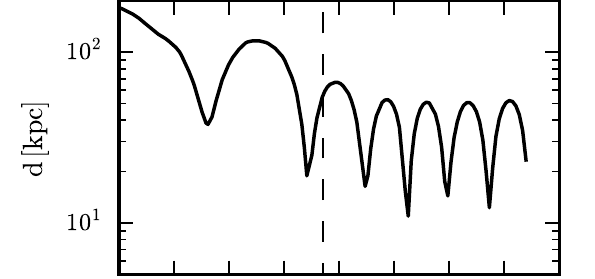}
\includegraphics[width=\columnwidth,trim={0 1.7cm 0 1.4cm}, clip]{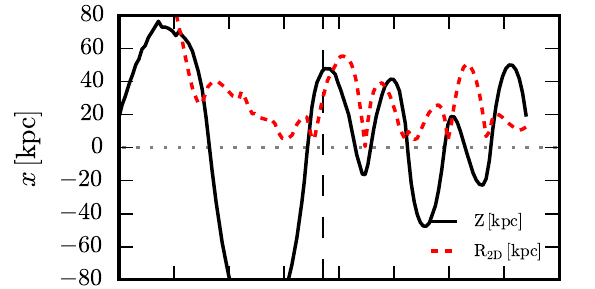}
\includegraphics[width=\columnwidth,trim={0 0.9cm 0 1.4cm}, clip]{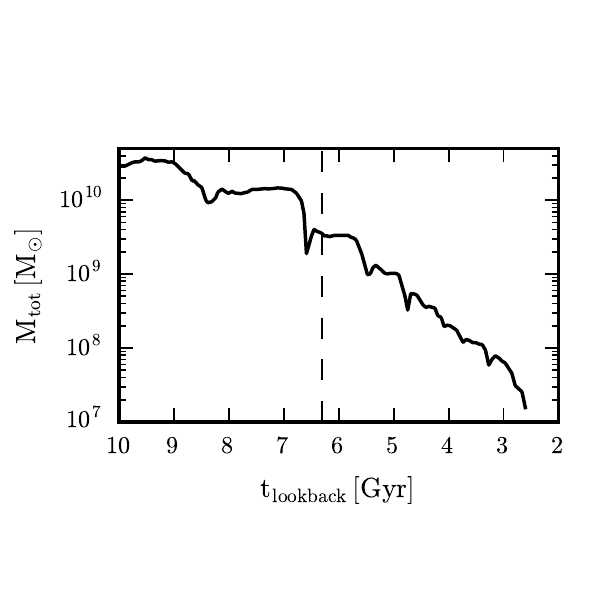}
\caption{The evolution of the distance (top panel), vertical height and cylindrical radius (middle panel), and total mass (bottom panel) of the satellite responsible for generating the dark matter wake. The vertical dashed line indicates the time at which dark matter-induced torque reaches its peak (see Fig.~\ref{dm:torq}).}
\label{sat}
\end{figure}

We now turn to the impact of satellites and dark matter. To calculate their effect, we follow the procedure of \citet{GWM15} which we briefly describe here for completeness. We define two rings with radii 14 kpc and 8 kpc, and, at every simulation snapshot, select disc particles within two separate galactocentric shells: $13.5 < r < 14.5$ kpc and $7.5 < r < 8.5$ kpc. We diagonalize the mass tensor associated with each of these particle subsets to obtain the orientation of the rings with respect to an inertial frame. For each case, the whole system is rotated such that a given ring's plane is aligned with the $X$–$Y$ plane. We then evenly sample 1000 positions along each ring and compute the torque on the ring from dark matter particles as

\begin{equation}
    \bm{\tau}_{\rm DM}^{\rm shell} = \sum _{i=1}^{1000} \mathbf{r_i} \times \mathbf{F}^{\rm shell}_i;
\end{equation}
where $\mathbf{r_i}$ represents the galactocentric distance vector to the $i$-th test particle along a ring and

\begin{equation}
    \mathbf{F}^{\rm shell}_i = \sum _{j=0}^{N_{\rm shell}} \mathbf{F}_{ij},
\end{equation}
is the gravitational force vector imparted on the $i$-th test particle by $N_{\rm shell}$ dark matter particles enclosed within a given spherical shell. For our purpose of identifying the source responsible for driving the phase spiral, we shall concern ourselves with the magnitude of the torque in the directions parallel to the ring's plane, $\tau _{\rm XY}$, (i.e. the $X$–$Y$ component of torque that can tilt the disc plane; the $Z$-component of the torque only affects the magnitude of the angular momentum). 

\begin{figure}
\includegraphics[scale=0.42,trim={0 1.8cm 0.6cm 0}, clip]{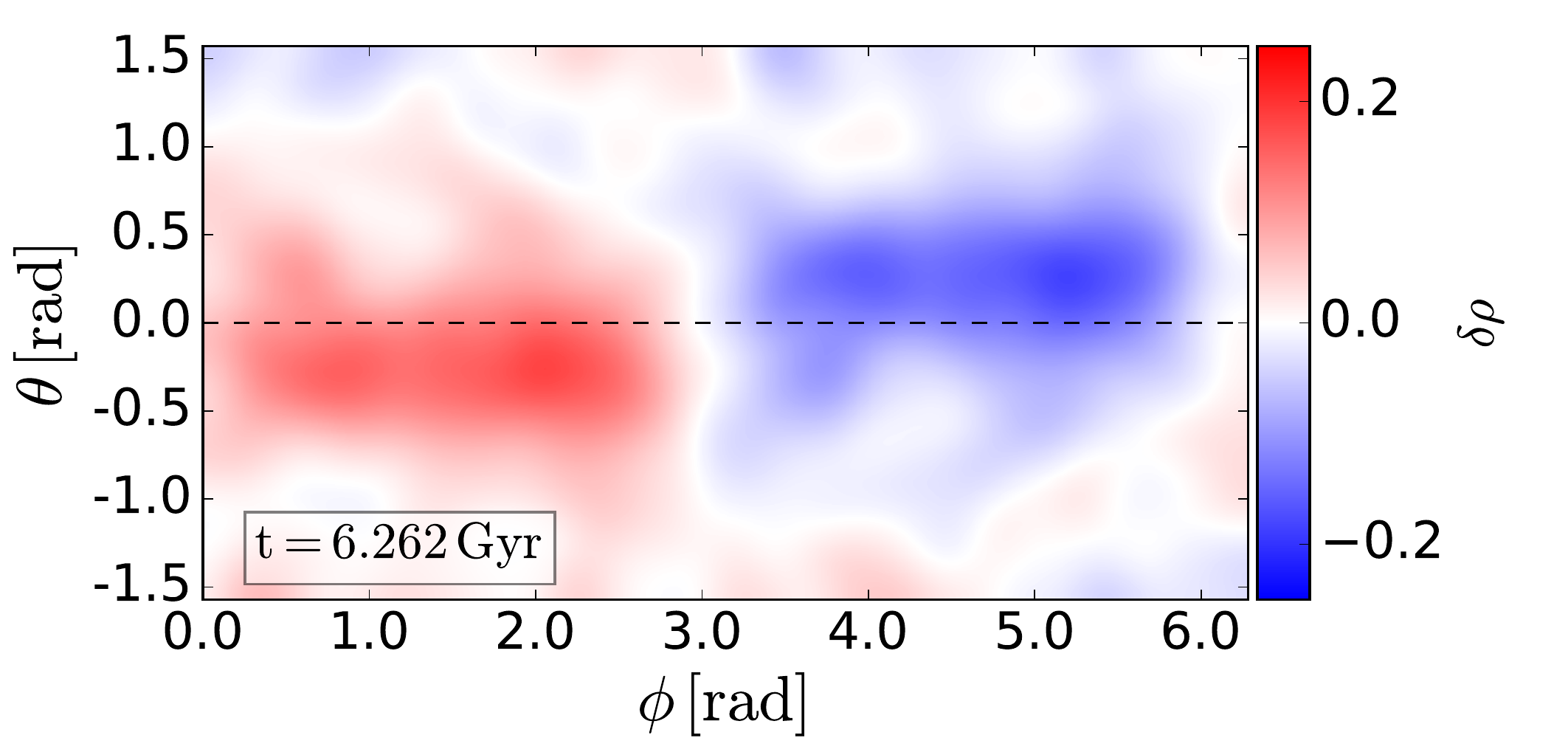}\\
\includegraphics[scale=0.42,trim={0 0 0.6cm 0}, clip]{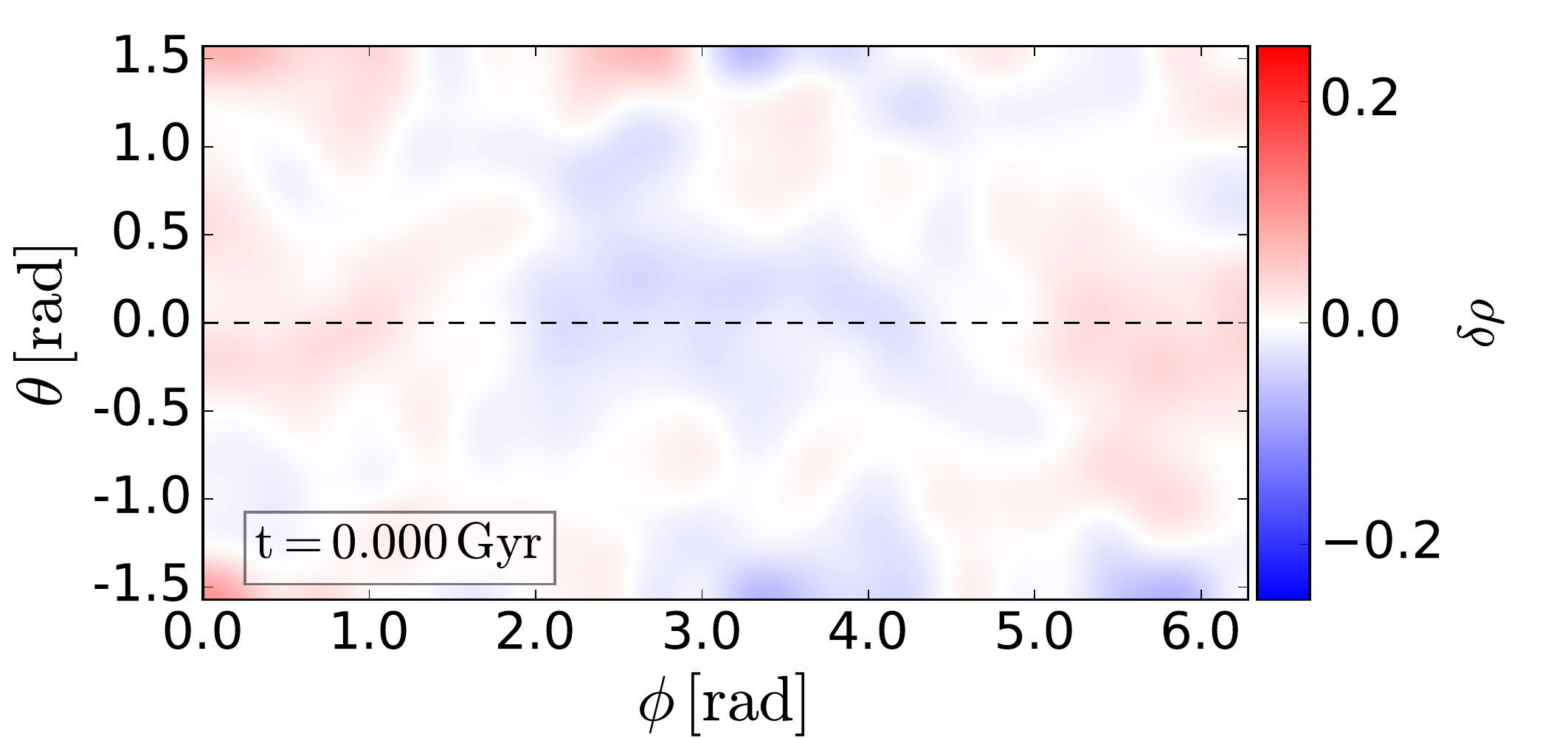}
\caption{Over-density maps, $\delta \rho$, obtained from the dark matter particles contained within a shell defined by spheres of 12 and 16 kpc galactocentric distances. These maps are obtained after rotating the original density by $\pi$ in $\phi$ and flipping about $\theta=0$, then subtracting this processed density map from the original in order to remove the quadrupolar triaxial halo feature. This enhances the dipolar signature of the wake, which is prominent at the time of maximum dark matter torque on the disc (upper panel) and inclined with respect to the disc midplane along $\theta = 0$ (dashed line). At the present day (lower panel), only a faint, patchy over-density map remains.}
\label{wake}
\end{figure}

The results are shown in Fig.~\ref{dm:torq} for the test particle rings of 8 kpc (left panel) and 14 kpc (right panel) radii, respectively. We focus on the time period spanning from $t_{\rm lookback} = 8$ Gyr to the present day because this period contains most of the evolution of the (thin) disc where phase spirals are expected to develop. The total torque from dark matter particles acting on each ring is clearly most dominant during the first half of this evolutionary period, and exhibits peaks approximately 6 Gyr ago. By subdividing the contributions from different spherical shells, we see that dark matter within 20 kpc of the galactic centre dominates the torque on the Solar ring, whereas the torque on the outer disc is dominated by material between 10 and 50 kpc. After $t_{\rm lookback} = 4$ Gyr, the torque is much diminished, although remains ever-present until the present day. In Appendix~\ref{appa}, we verify that this torque is numerically well-converged.

The circular symbols in Fig.~\ref{dm:torq} depict the torque imparted on each ring by satellite galaxies (multiplied by a factor of 4 to aid comparison to the dark matter-induced torque) and are coloured according to their total mass. Note that the spike in torque imparted by a satellite of $\sim 10^{10}$ $\rm M_{\odot}$ at around the look back time of 6.5 Gyr precedes the peak in dark matter torque by order $\sim 100$~Myr. Fig.~\ref{sat} shows the evolution of the galactocentric distance (top panel), vertical height and cylindical radius (middle panel), and the total mass (bottom panel) of this satellite. Interestingly, the orbit and mass-loss history of this satellite up to the third or fourth pericentric passage appears qualitatively consistent with some models of Sgr \citep[e.g.][]{Vasliev_Belokurov2020,Huntetal2021}, albeit occurring some gigayears prior to the real Sgr. Its mass is $3\times 10^{10}$ $\rm M_{\odot}$ prior to infall at $t_{\rm lookback}\sim 8.5$ Gyr. The next pericentric passage occurs $\sim 2$ Gyr later on a near-polar orbit with a relative vertical velocity of $\sim 400$ $\rm km\,s^{-1}$, and corresponds to the time at which peak satellite-induced torque is attained. Incidentally, this rapid encounter correlates with the onset of the local two-armed phase spiral feature shown in the lower panel of the second column in Fig.~\ref{collage}: likely a manifestation of satellite-induced ``breathing modes''\footnote{A breathing mode manifests as a two-armed spiral in the $Z$-$V_Z$ plane as the disc vertically expands and contracts symmetrically about the midplane, contrary to the one-armed bending mode where the disc is locally displaced as a whole into an asymmetric oscillation about the midplane.} \citep[as described in e.g.][]{WBC14,Hunt2022}. Subsequent pericentric passages of this satellite register decreasing values of $\tau _{\rm XY}$, particularly on the outer disc ring, as its mass decreases owing to tidal stripping.

An important feature seen in (particularly the right panel of) Fig.~\ref{dm:torq} is that peaks in dark matter torque either coincide with (or follow shortly after) peaks of satellite-induced torque. This is a smoking-gun signature of dark matter halo wakes that form behind satellite galaxies and amplify to produce a gravitational torque on the disc greater than that of the satellite behind which it formed. To show that a dark matter wake forms, we calculate an all-sky map of the dark matter over-density in a spherical shell between galactocentric radii of 12 and 16 kpc by performing the following steps \citep[see][for a more thorough description]{GWM15}:

\begin{itemize}
    \item we apply a smoothing kernel to the dark matter particle distribution to estimate the density for each particle;
    \item we transform dark matter particle coordinates into spherical polar coordinates centred on the potential minimum of the galaxy, and for each particle in the spherical shell, calculate a polar grid of densities; 
    \item we calculate an average density for the spherical shell to find the normalised dark matter over-density at each grid point:
    \begin{equation}
        \hat{\rho} = \frac{\rho _{\rm grid} (R,\phi,\theta) - \bar{\rho}_{\rm shell}}{\bar{\rho}_{\rm shell}}
    \end{equation}
    \item to eliminate the triaxial halo signal from the map, we rotate the map by $\pi$ in $\phi$ and flip along the $\theta=0$ axis to produce $\hat{\rho}_{\rm flip} (R,\phi,\theta)$, then calculate dipolar wake over-density as
    \begin{equation}
        \delta \rho = \frac{\hat{\rho} - \hat{\rho}_{\rm flip}}{2}.
    \end{equation}
\end{itemize}

To better visualise the all-sky map, we perform the procedure above on a re-simulation of the original halo Au 6 with a factor 8 higher dark matter mass resolution (see Appendix~\ref{appa} for more details). The all-sky map for $\delta \rho$ at the time the dark matter-imparted torque reaches a maximum is shown in the upper panel of Fig.~\ref{wake}. The clear dipole observed is the signature of the wake, the major axis of which is misaligned with the $\theta = 0$ vector that defines the mid-plane of the galactic disc. As shown by \citet{GWG16}, a dark matter wake with a similar such alignment creates gravitational forces that act vertically on the disc to deform it into a warp-like pattern. The lower panel of Fig.~\ref{wake} shows the all-sky map for $\delta \rho$ at $z=0$: the wake has clearly decayed and all that remains is a relatively weak (and somewhat patchy) over-density. This correlates with the sustained low and roughly constant torque at values $\sim 5$ times lower than the peak wake activity seen in the last few Gyr of evolution in Fig.~\ref{dm:torq}, but still dominates over satellite-induced torque at all times. The phase spiral is left to wind-up and decay during this late epoch into the comparatively weak pattern seen in Fig.~\ref{multi:now}.

\section{Conclusions and Discussion}
\label{conclusions}

The {\it Gaia} phase spiral \citep{Antojaetal2018}, in addition to several other dynamical features \citep[e.g. planar radial motions:][]{Kawata2018}, observed in the Milky Way is a sign that our Galaxy is in dynamical disequilibrium. The nature of the perturbation that set it in motion is the subject of much debate. Most of the theoretical work on the phase spiral has involved either very simplified toy models \citep{Binney_Schoenrich2018} or idealised $N$-body simulations. Such numerical experiments sacrifice complex physics and cosmological environment for high resolution and/or a high-degree of control over the setup, and have provided several possible mechanisms including a buckling bar \citep{Khoperskovetal2019} and the recent passages of Sgr \citep{LMJ19,Huntetal2021}. 

In this paper, we have studied one of the first \textlcsc{Superstars} cosmological magnetohydrodynamical simulations containing $\sim 10^8$ disc star particles at $z=0$. This provides a complementary view of the problem by connecting it to a wide range of dynamical phenomena inherent to galaxy formation in the $\Lambda$CDM cosmological paradigm, such as continued gas accretion \& star formation, satellite interactions, and feedback. Our main conclusions are as follows:

\begin{itemize}
    \item At late times ($\rm t_{lookback} \lesssim 3$ Gyr) including the present day, our simulation shows a range of phase spiral features at multiple radii and azimuths over the stellar disc. At Solar-like positions, phase spirals have typical over-density amplitudes between $\sim 0.05$ and $0.15$ and radial velocity amplitudes of $\sim10$ $\rm km\, s^{-1}$. Remarkably, these properties are not unlike those of the {\it Gaia} snail shell despite all the complexities of the simulation and its differences with respect to the Milky Way. Our simulated phase spirals have 2-3 wraps in dimensionless coordinates - similar to the observed dimensionless phase spiral of \citet{Hunt2022} - but is more extended than observed in $Z$-$V_Z$ coordinates.
    \item The phase spirals are most clearly visible for coeval stellar populations of intermediate age (2-6 Gyr old), whereas older stars have a comparatively weak signal. 
    \item For star particles located in the outer disc ($R\sim 14$ kpc), phase spirals exhibit the same trends with age as for the Solar-like populations, but are ``squashed'' along the $V_Z$-axis owing to the lower disc surface density and vertical restoring force \citep[in agreement with][]{LMJ19,Garcia-Condeetal2022}. They are also less-tightly wound than their Solar-position counterparts owing to their longer dynamical timescales, which indicates the outer Galactic disc is a promising place to look for signatures of past perturbations.
    \item We present new insights into a scenario for the formation of the phase spiral: first, a satellite of total infall mass $\sim 10^{10}$ $\rm M_{\odot}$ generates the formation of a wake over-density in the dark matter halo \citep[see also][]{GWM15} during its first pericentric passage $\rm t_{lookback} \sim 6$-$7$ Gyr. This dark matter wake generates a strong gravitational torque parallel to the disc; it is approximately 8 times as strong as the direct torque imparted by the satellite at Solar radii. As a result, a strong warp forms in the disc which evolves into a global corrugation pattern. Locally, the oscillations associated to the corrugation pattern wrap up into spirals in the vertical phase plane. 
    \item In our simulation, phase spirals are ever-present: they first appear during the early epochs of disc formation/evolution under the action of the dark matter halo wake, and are sustained until redshift zero, long after the main peak of the wake decays. The precise mechanism of this survival is unclear. However, idealised simulations have shown that the self-gravity of discs can maintain generations of bending waves for significant periods of time \citep[e.g.][]{Chequers2017,Chequers2018}. We speculate that this mechanism occurs in our cosmological simulation as well, and defer a dedicated investigation to a separate future study.
\end{itemize}

Our findings have significance for the Sgr interpretation of the phase spiral: by fortuitous circumstance, the wake-inducing satellite shares many similar properties to the inferred orbit and mass-loss history of some dynamical models of Sgr \citep[e.g][]{Vasliev_Belokurov2020}. The main difference is that, in our simulation, its orbit is offset several gigayears into the past with respect to the real Sgr. A crude accounting of this offset would correspond to a ``present day'' satellite mass of $\sim 3\times 10^8$ $\rm M_{\odot}$, and a phase spiral like the one shown in the lower-right panel of Fig.~\ref{collage} which would also be visible in the youngest stellar populations; in better qualitative accord with the age trends of observations \citep[e.g.][]{LMJ19,Bland-Hawthorn2019}. Thus, it seems plausible that a dark matter wake associated with the initial passages of Sgr could be playing a major role in the formation and propagation of the {\it Gaia} snail shell. This contrasts somewhat with the findings of recent idealised $N$-body simulations: for example, \citet{Laporte2018,LMJ19} showed that their Sgr analogue created a dark matter wake that dominated the torque on its first pericentric passage, but then subsequently decayed to a negligible level before direct torques from Sgr ``reset'' the phase spiral pattern. It is possible that the properties of Milky Way-mass halos and their satellite distributions in $\Lambda$CDM cosmological simulations alter the halo response of wake amplification processes \citep{Vesperini-Weinberg2000, GWM15} relative to idealised setups comprising a single object on a prescribed orbit in a smooth, spherical halo.

Importantly, a non-negligible role of a dark matter wake in the formation of the phase spiral could complicate the mapping between the phase spiral properties and those of Sgr. Indeed, the $\sim 10^{10}$ $\rm M_{\odot}$ infall mass of our wake-inducing satellite is somewhat less massive than that favoured by some idealised models \citep{LMJ19,BHTC2021}, but may help alleviate some of the discrepancies between the mass of the Sgr remnant and phase spiral properties discussed by \citet[][]{Bennettetal2022}. 

In addition to the mechanisms discussed in this paper,  \citet{Tremaine2023} used a simple model to suggest a separate collective response of many ``Gaussian noise'' perturbations from subhalos and/or giant molecular clouds can reproduce many properties of the phase spiral. While some of these types of perturbation should exist in our simulation, it is difficult to assess their prevalence and impact on the formation and destruction of the phase spiral alongside the other effects present. Controlled numerical simulations with prescribed spectra of small-scale perturbers would help clarify the situation.

With respect to other cosmological simulations, \citet{Garcia-Condeetal2022} found that pericentric passages of a satellite of infall mass similar to that of our Sgr-like satellite correlated with the emergence of phase spirals in their simulation. However, they remark that the pericentre of this satellite is larger than that of Sgr and conclude that it is unlikely to be the sole contibutor to the perturbation responsible for their phase spirals. We speculate that the dark matter halo wake mechanism discussed in this work is present also in their simulation.

Finally, we remark that the dynamical response of the disc to perturbations has a dependency on the kinematics of newborn coeval stellar populations which in turn depends on our galaxy formation model as well as the precise assembly history of the simulated galaxy. For example, it is possible that the phase space coordinates of star particles older than $\sim 6$ Gyr just prior to the wake perturbation are different (perhaps dynamically hotter) than those of the Milky Way, which may explain the lack of clear signal for these star particles in our simulation. We note also that our simulation does not appear to produce the recently discovered two-armed phase spiral in the inner disc \citet{Hunt2022}; however, this may not be surprising because its origin is speculatively linked to bar/spiral structure, whereas our simulation does not contain a bar according to typical definitions \citep[e.g.][]{Athanassoula2002,Algorry2017,Fragkoudi+Grand+Pakmor+19}. Thus many questions remain open. Nevertheless, our results underline the difficulty in interpreting the complex dynamical history of the Galaxy, and expose a new link between phase spiral features like the {\it Gaia} Snail shell and the dark matter distribution around the Galaxy. To make further progress on the phase spiral, future cosmological simulations with calibrated Sgr analogues will need to be performed. More generally, our study highlights the potential for a suite of \textlcsc{Superstars} simulations (Fragkoudi et al. in prep) to scrutinise galactic dynamics in the cosmological context, such as the nature of galactic spiral arms and bar formation and evolution. We defer these tasks to future work.

\section*{Acknowledgements}
The authors thank the referee for a prompt, constructive report. RG acknowledges financial support from the Spanish Ministry of Science and Innovation (MICINN) through the Spanish State Research Agency, under the Severo Ochoa Program 2020-2023 (CEX2019-000920-S), and support from an STFC Ernest Rutherford Fellowship (ST/W003643/1). FAG acknowledges support from ANID FONDECYT Regular 1211370, the Max Planck Society through a “Partner Group” grant and ANID Basal Project FB210003. FvdV is supported by a Royal Society University Research Fellowship (URF\textbackslash R1\textbackslash 191703). The authors gratefully acknowledge the Gauss Centre for Supercomputing e.V. (www.gauss-centre.eu) for funding this project by providing computing time on the GCS Supercomputer SUPERMUC-NG at Leibniz Supercomputing Centre (www.lrz.de).

\section*{Data Availability}
High level data underlying this article will be shared on reasonable request to the corresponding author.

\bibliographystyle{mnras}
\bibliography{main.bbl}

\appendix

\section{Phase spirals at the present day}
\label{appa2}

In this appendix, we present several versions of the phase spirals depicted in Fig.~\ref{multi:now} in order to demonstrate how they depend on numerical resolution and the sample volume. We also show the dimensional counterpart for completeness. 

\begin{figure*}
\includegraphics[scale=0.68,trim={0 0 0 0}, clip]{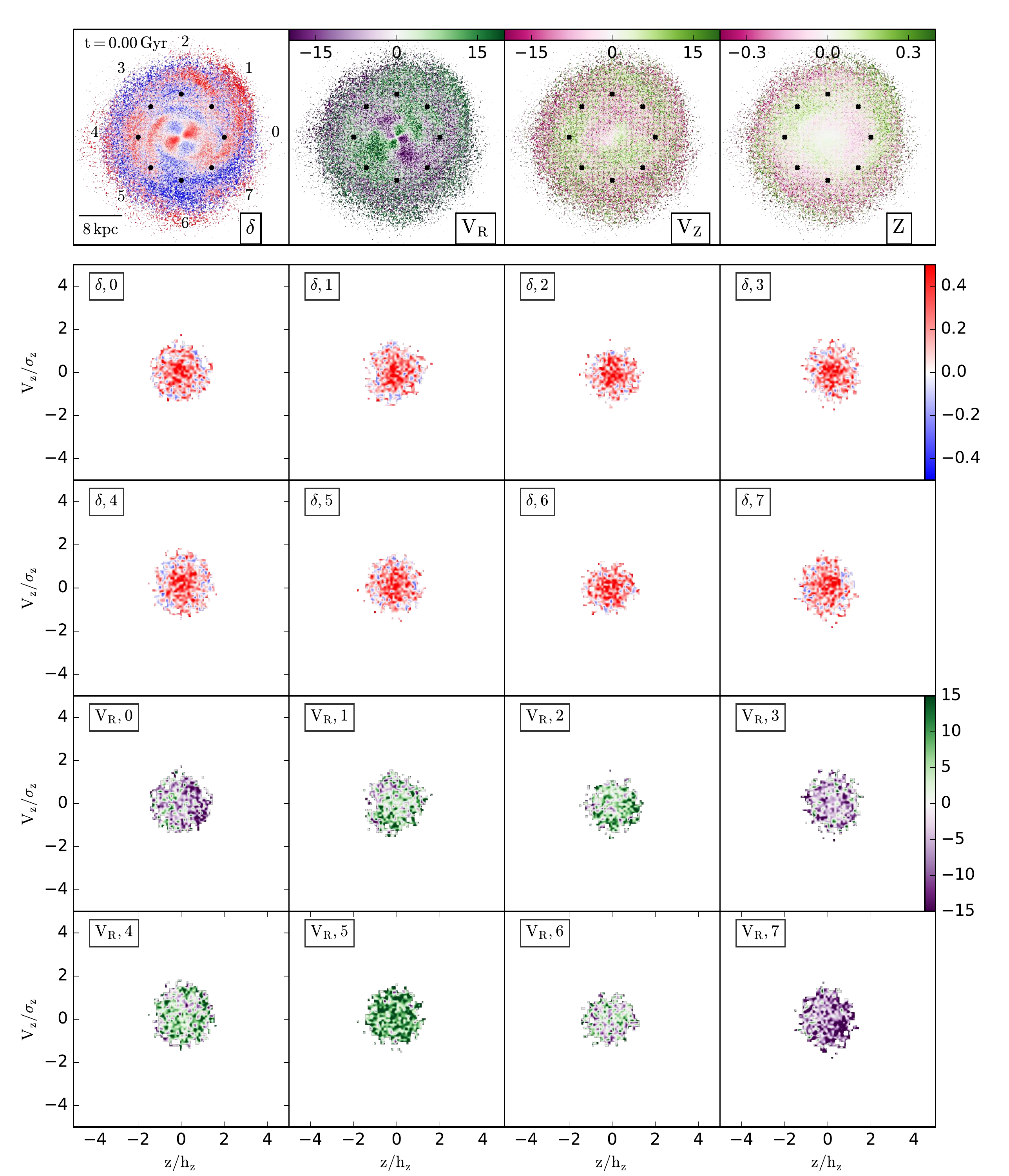}
\caption{As Fig.~\ref{multi:now}, but for a simulation of the same halo at level 4 resolution: star particle mass of $\sim 5 \times 10^4$ $\rm M_{\odot}$.}
\label{multi:lowres}
\end{figure*}

Fig.~\ref{multi:lowres} shows a version of Fig.~\ref{multi:now} for the level 4 resolution simulation of the same halo. It is evident that phase spirals are not resolved at the present day at this resolution. This demonstrates the ability of our \textlcsc{superstars} simulations to capture otherwise unresolved galactic dynamical structures.

\begin{figure*}
\includegraphics[scale=0.68,trim={0 0 0 0}, clip]{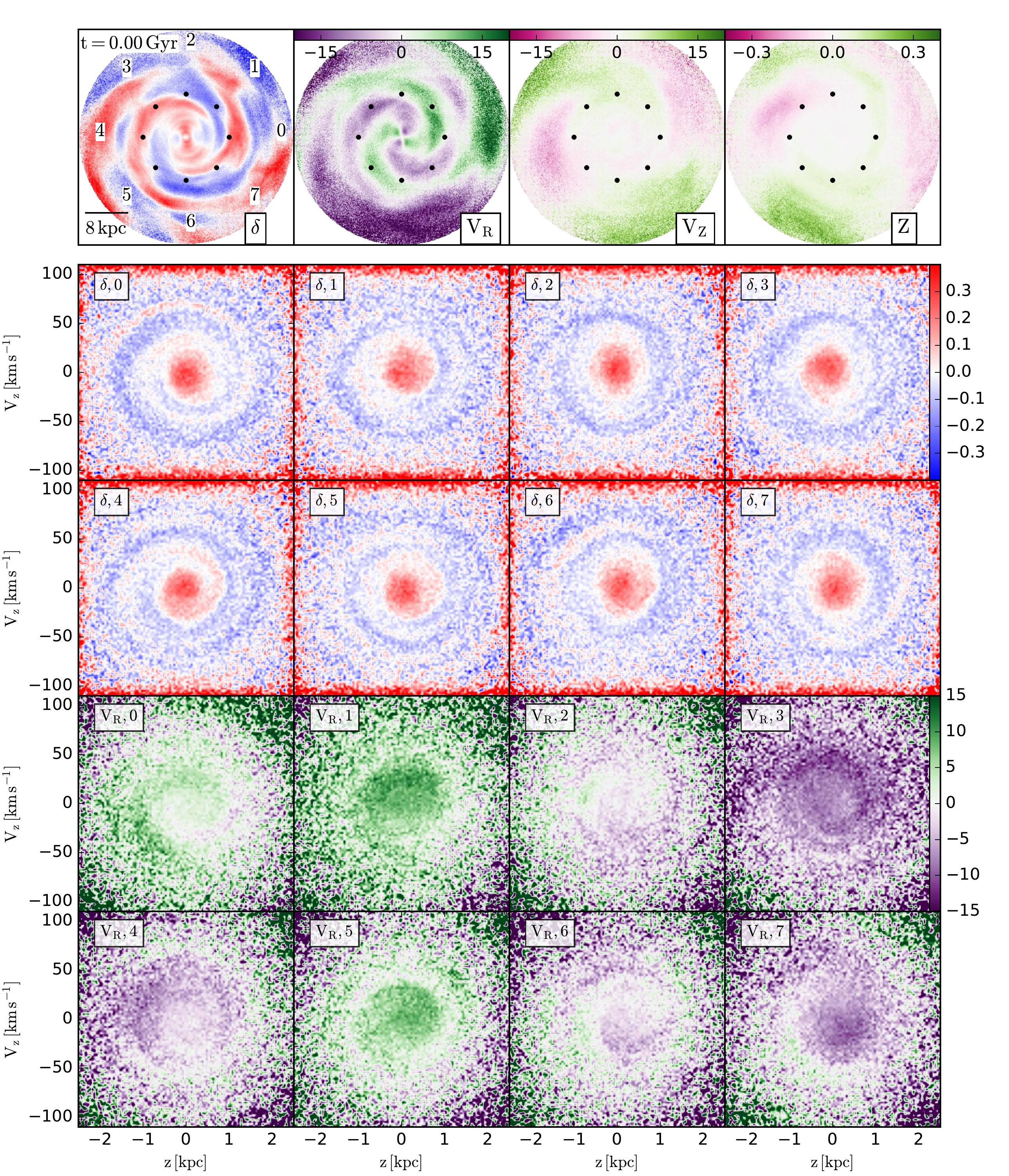}
\caption{As Fig.~\ref{multi:now}, but the phase spirals are shown in the dimensional $Z$-$V_Z$ plane. The amplitudes of both the over-density (second and third rows) and the radial velocity (fourth and fifth rows) are the same as the dimensionless phase spirals shown in Fig.~\ref{multi:now}.}
\label{multi:realdims}
\end{figure*}

\begin{figure*}
\includegraphics[scale=0.68,trim={0 0 0 0}, clip]{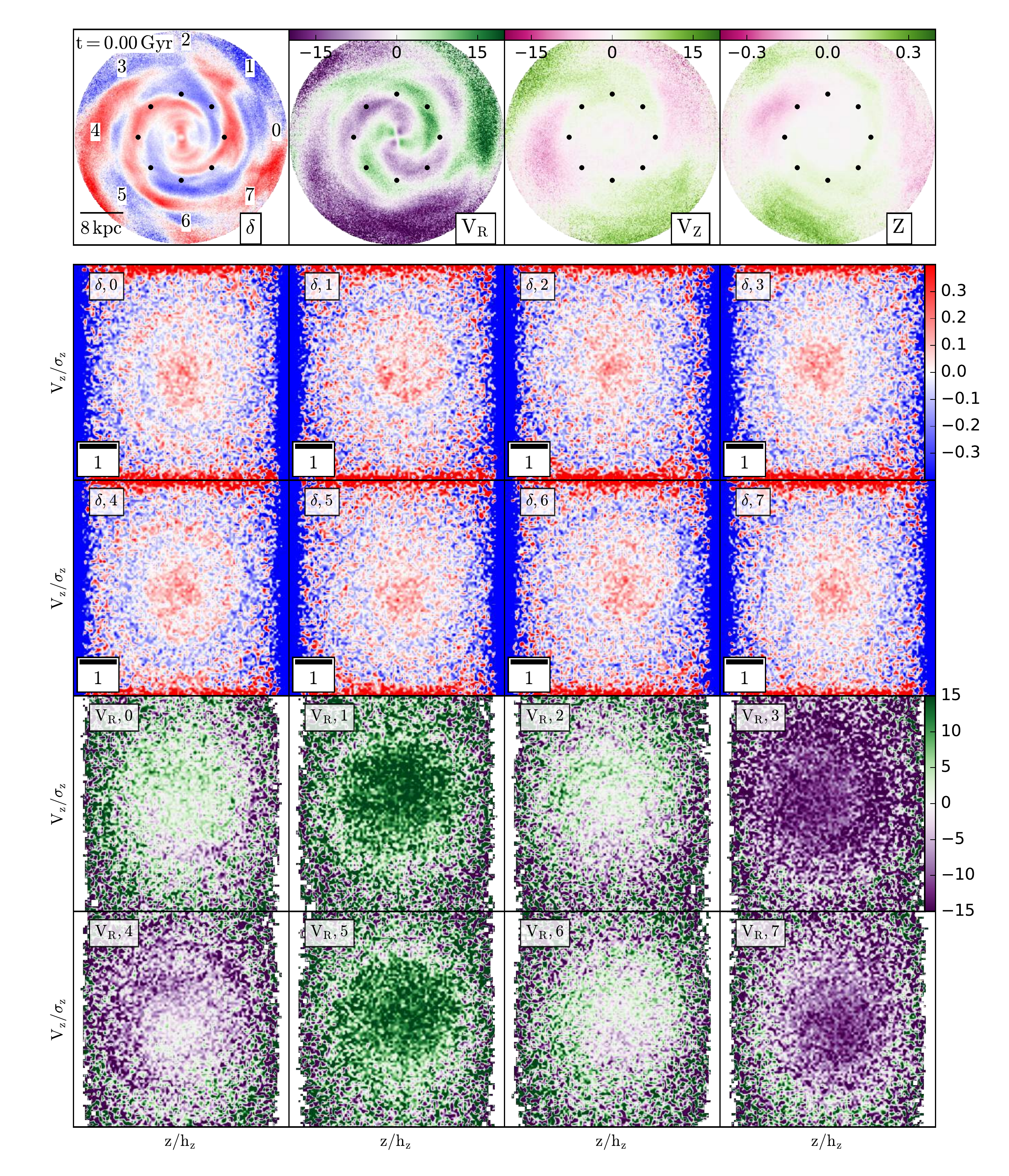}
\caption{As Fig.~\ref{multi:now}, but for star particles selected in spheres of 1.5 kpc radii at Solar-like positions. The range of the dimensionless axes have been shortened to $\pm 3$.}
\label{multi:smallvol}
\end{figure*}

Fig.~\ref{multi:realdims} shows the phase spirals without normalising $Z$ and $V_Z$ by $h_Z$ and $\sigma _Z$, respectively. In terms of both over-density and radial velocity, the amplitudes of each phase spiral are essentially identical to the dimensionless phase spirals of Fig.~\ref{multi:now}. The phase spirals seem to extend to larger heights and velocities compared to the the {\it Gaia} snail shell \citep[e.g.][]{LMJ19,Antoja2023}. The reason for this difference is unclear given that our simulation is not tailored to the Milky Way. We note also that our sample volume is larger than those typically probed by {\it Gaia}. Unfortunately, although the phase spirals for star particles selected from spheres of 1.5 kpc radius are discernible (see Fig.~\ref{multi:smallvol}), they are evidently noisier and less clear than the 3 kpc volume shown in Fig.~\ref{multi:now}. We therefore opt to show the phase spirals clearly for 3 kpc spherical volumes throughout this paper.

\section{Resolution dependence of the dark matter wake}
\label{appa}

\begin{figure}
\includegraphics[width=\columnwidth,trim={0 0 0.5cm 0}, clip]{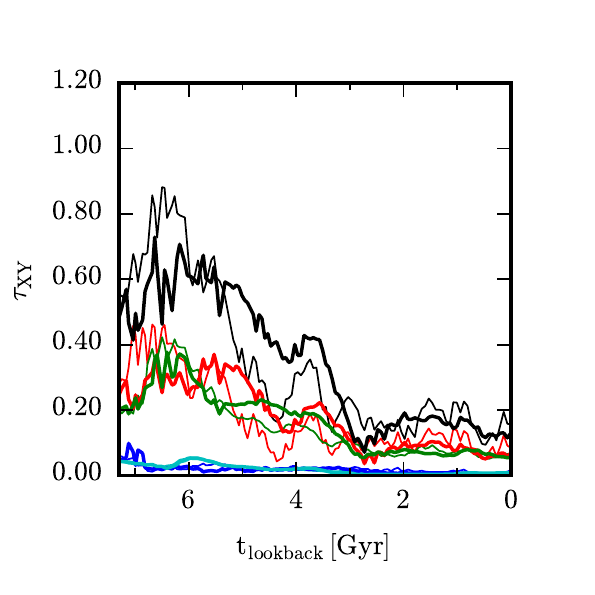}
\caption{As the right panel of Fig.~\ref{dm:torq}, but for the dark matter torques from a re-simulation with 8 times better dark matter resolution (denoted DM8) as well as the fiducial simulation (denoted Superstars64). Each thin curve traces its thick curve counterpart very well, which indicates a high level of convergence for dark matter induced torque.}
\label{dmres}
\end{figure}

Because dark matter wakes develop over large volumes in the halo, it is logical to ask whether the dark matter particle resolution is high enough (and noise low enough) to capture the dynamics \citep[e.g.][]{Garavito-Camargo2019}. To demonstrate that the torques induced by the dark matter wake are robust to resolution changes, Fig.~\ref{dmres} shows a reproduction of the right hand panel of Fig.~\ref{wake} for the \textlcsc{Superstars} simulation (here denoted ``Superstars64'') as well as a re-simulation of the same system, ``DM8'' with 8 times as many dark matter particles (but with the standard single star particle formed per gas cell, i.e., ``level 4'' stellar resolution). It is evident that, for each radial shell considered, the salient features of the evolution of the torque acting perpendicular to the disc minor axis are preserved at both resolution levels. The main appreciable difference is that the strength of the torque decays more slowly after the peak at $\sim 6$ Gyr for the DM8 run compared to the Superstars64 run. However, the difference is slight and could potentially be accounted for by minor stochastic variations in, for example, the precise infall time, mass, and orbit of the wake-generating satellite; such variations can arise from the ``butterfly effect'' phenomenon \citep[e.g.][]{Genel2019,GMP21} or resolution changes (although we note that the evolution of the putative satellite in the Superstars64 and DM8 simulations is almost identical). Nevertheless, the convergence seen in Fig.~\ref{dmres} indicates that the torque imparted by the dark matter wake is captured in our \textlcsc{Superstars} simulation. 

\bsp	
\label{lastpage}
\end{document}